\documentclass[a4paper,11pt]{article}
\usepackage{jinstpub} 
\usepackage{natbib}
\usepackage{lineno}
\usepackage{graphicx}
\usepackage{caption}
\usepackage{subfig}
\usepackage{multirow}
\usepackage{booktabs}
\usepackage{indentfirst} 
\usepackage{booktabs}
\usepackage{siunitx}

\title{\boldmath Thermal Cycling Reliability of Hybrid Pixel Sensor Modules for The ATLAS High Granularity Timing Detector}

\author[1,2]{Y.~Li,}
\author[3]{A.~Aboulhorma,}
\author[3]{M.~Ait~Tamlihat,}
\author[4,5]{H.M.~Alfanda,}
\author[6]{N.~Atanov,}
\author[6]{O.~Atanova,}
\author[7]{I.~Azzouzi,}
\author[1]{J.~Barreiro~Guimar\~{a}es~Da~Costa,}
\author[8]{T.~Beau,}
\author[9]{D.~Benchekroun,}
% ...
\affiliation[1]{Institute of High Energy Physics, Chinese Academy of Sciences, Beijing, China}
\affiliation[2]{University of Chinese Academy of Sciences (UCAS), Beijing, China}
\affiliation[3]{Facult\'e des sciences, Universit\'e Mohammed V, Rabat, Morocco}
\affiliation[4]{State Key Laboratory of Dark Matter Physics, Tsung-Dao Lee Institute, Shanghai Jiao Tong University, Shanghai, China}
\affiliation[5]{State Key Laboratory of Dark Matter Physics, School of Physics and Astronomy, Shanghai Jiao Tong University, Key Laboratory for Particle Astrophysics and Cosmology (MOE), SKLPPC, Shanghai, China}
\affiliation[6]{Affiliated with an international laboratory covered by a cooperation agreement with CERN}
\affiliation[7]{Moroccan Foundation for Advanced Science Innovation and Research (MAScIR), Rabat, Morocco}
\affiliation[8]{LPNHE, Sorbonne Universit\'e, Universit\'e Paris Cit\'e, CNRS/IN2P3, Paris, France}
\affiliation[9]{Facult\'e des Sciences Ain Chock, Universit\'e Hassan II de Casablanca, Casablanca, Morocco}
% ...

\emailAdd{ylli@ihep.ac.cn}

\abstract{The reliability of bump connection structures has become a critical aspect of future silicon detectors for particle physics. The High Granularity Timing Detector (HGTD) for the ATLAS experiment at the High-Luminosity Large Hadron Collider will require 8032 hybrid pixel sensor modules, composed of two Low Gain Avalanche Diode sensors bump-bonded to two readout ASICs and glued to a passive PCB\@. The detector will operate at low temperature (\SI{-30}{\degreeCelsius}) to mitigate the impact of irradiation. 
The thermomechanical reliability of flip-chip bump connections in HGTD modules is a critical concern, particularly due to their characteristically lower bump density (pixel pitch dimensions of \SI{1.3}{\milli\meter} $\times$ \SI{1.3}{\milli\meter}). This paper elaborates on the challenges arising from this design characteristic. Finite element analysis and experimental testing were employed to investigate failure modes in the flip-chip bump structures under thermal cycling from \SI{-45}{\degreeCelsius} to \SI{40}{\degreeCelsius} and to guide the module redesign. 
The optimized design demonstrates significantly enhanced robustness and is projected to fulfill the full lifetime requirements of the HGTD.}


\begin{document}
\maketitle
\flushbottom

\section{Introduction}

Progress in particle physics is dependent on advances in detector technology. 
Among these, charged particle detectors are paramount, with semiconductor-based devices now providing the finest spatial resolution.
Hybrid Pixel Sensors (HPS) have been the leading technology for tracking and vertexing in high-energy collision experiments due to their superior performance~\cite{particle_detector}.
They constitute the core detection system in major experiments, including CMS~\cite{CMS}, ATLAS~\cite{ITk-TDR,strip}, and BELLE-$\mathrm{\uppercase\expandafter{\romannumeral 2}}$\cite{belle}.
    
The Large Hadron Collider (LHC), the largest scientific instrument ever built, is scheduled for an upgrade to its high-luminosity (HL) phase, commencing in 2030 and lasting over a decade. The HL-LHC will deliver unprecedented luminosity, resulting in a high number of simultaneous proton–proton interactions, a phenomenon known as pile-up. This significantly complicates particle identification and tracking. To mitigate the effects of pile-up in the ATLAS experiment, a novel detector, the High Granularity Timing Detector (HGTD)~\cite{TDR}, is under construction to complement a new, enlarged silicon Inner Tracker (ITk)~\cite{ITk-TDR} by providing high-precision timing information. This will mark the first deployment of a timing detector based on Low Gain Avalanche Diode (LGAD) technology at a major collider experiment, while similar technology is being developed for the CMS experiment~\cite{MIP}.

\begin{figure}[h]
        \centering
        \subfloat[\label{fig:HGTD}]{\includegraphics[width = 0.45\textwidth]{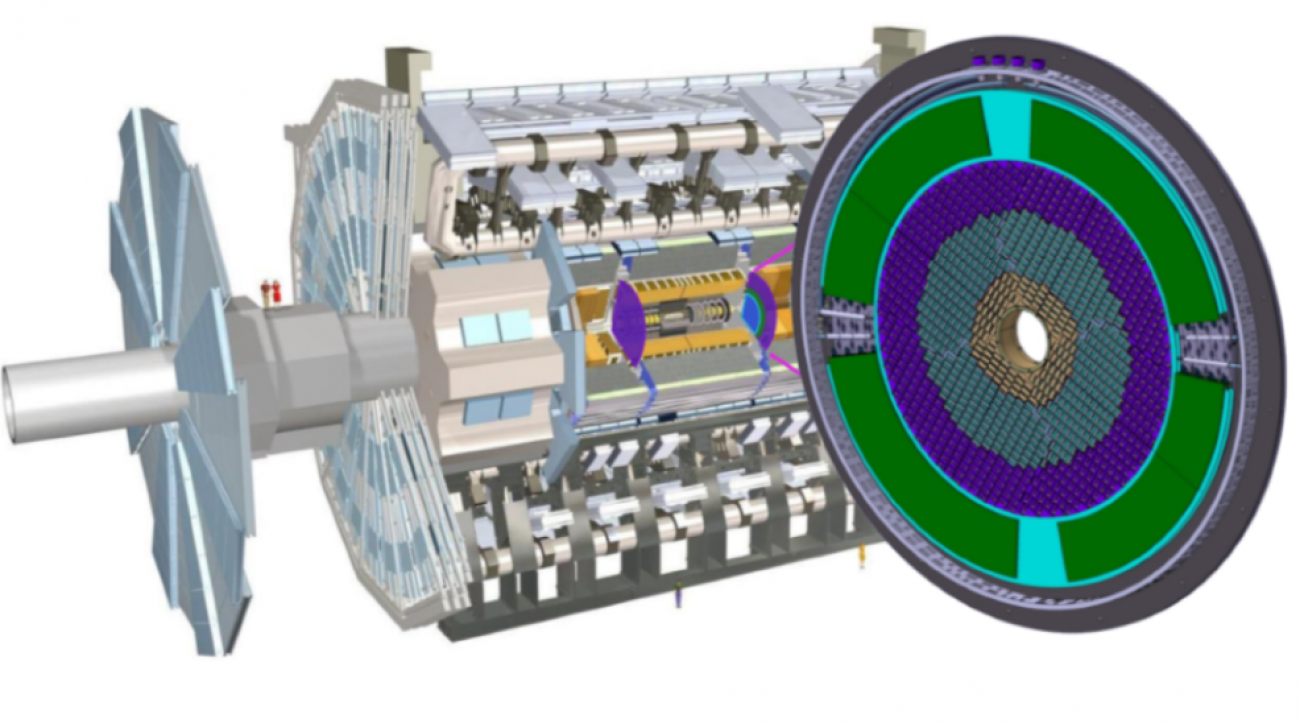}}
        \hfill
        \subfloat[\label{module} ]{\includegraphics[width = 0.45\textwidth]{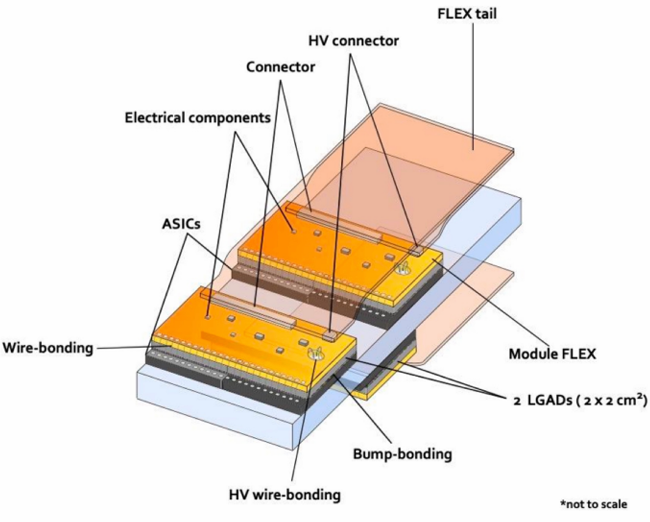}}
        \hfill
    \caption{The High Granularity Timing Detector. (a) The HGTD is composed of two disks placed on each side of the ITk. (b) The HGTD module approximate dimensions are $\mathrm{2cm\times4cm\times2mm}$. The module consists of two readout ASICs bump-bonded to two LGAD sensors, glued together onto a PCB that provides power and communication for the ASICs through wire-bonds. The FLEX tail connects the modules to the DAQ system.}
    \end{figure}

HPS are utilized in the HGTD\@. Their hybrid architecture, where the sensing element and the readout electronics are fabricated on separate substrates and later interconnected, offers flexibility in optimizing the radiation-sensing element for peak performance. An HGTD module comprises two readout Application-Specific Integrated Circuits (ASICs) bump-bonded to two LGAD sensors, which are glued onto a PCB, as illustrated in Figure~\ref{module}.
Each sensor contains a $15 \times 15$ array of pixels with a pitch of \SI{1.3}{\milli\meter} $\times$ \SI{1.3}{\milli\meter} and an active thickness of \SI{50}{\micro\meter}. Sensor wafers with a thickness of \SI{775}{\micro\meter} are employed in the HGTD, while ASIC wafers have a thickness of \SI{300}{\micro\meter}. To reduce the module thickness, the initial design applied a thinned 300-\SI{}{\micro\meter} sensor and 300-\SI{}{\micro\meter} ASIC\@. A total of 8032 modules will be deployed in the HGTD\@.
Despite their advantages, the manufacture of these modules is challenging. Their stringent precision requirements present difficulties for mass production. Furthermore, the harsh irradiation environment (a maximum fluence of $\mathrm{2.5\times10^{15}~n_{eq}cm^{-2}}$) and required long operational lifetime necessitate exceptional device stability and durability.

The hybridization process begins with the deposition of under-bump metallization (UBM) layers on both the LGAD sensor and readout ASIC wafers. Solder bumps are then formed on the ASIC UBM layer. Following these depositions, both wafers are diced into individual tiles.
Final hybridization is accomplished via flip-chip bonding, wherein a single ASIC tile is precisely aligned and bonded to a corresponding sensor tile. This technique applies controlled heat and pressure to form reliable electrical and mechanical interconnections between the substrates. The resulting structure of an ASIC bump-bonded to an LGAD sensor is termed a hybrid.
Later, two hybrids are adhesively mounted onto a PCB using either a pick-and-place machine or jig tools. The assembly is completed by wire-bonding the ASICs to the PCB\@. Adhesion between the PCB and the hybrids is achieved using an epoxy-based glue, Araldite 2011.
To mitigate irradiation effects, the modules operate at a low temperature of \SI{-30}{\degreeCelsius}. Considering detector commissioning, module replacements and power-off during operation, guaranteeing thermomechanical reliability against temperature variations is essential to ensure survival throughout the full HGTD operational lifetime.  Each hybrid contains 225 solder bumps for the pixels, with additional bumps located on the guard ring to ensure mechanical stability.

Thermal cycling tests on initial production modules revealed widespread bump connection failures after only a few cycles. These modules consisted of LGAD sensors thinned to \SI{300}{\micro\meter} and hybrids incorporating 30 additional bumps on the guard ring. The flip-chip bumps connecting the sensor and ASIC failed through either disconnection or fracture.These results motivated a detailed study of the HGTD modules and the development of solutions to enhance their thermal reliability. While flip-chip bonding reliability under thermal cycling has been extensively studied~\cite{flip-chip1,flip-chip2}, primarily using the finite element method (FEM), the HGTD presents a unique challenge. 
A major cause of such failures is the mismatch in the coefficient of thermal expansion (CTE) among different materials. This mismatch induces reciprocating shear stress and strain in the solder joints during thermal cycles, ultimately leading to fatigue failure.
In HGTD modules, the \SI{1.3}{\milli\meter} pitch between adjacent bumps is relatively large compared with the bump dimension, leading to a lower bump density than in common electronic packaging (e.g BGA). This lower bump density reduces the structural integrity and poses a greater challenge to mechanical stability.
To address this issue, this paper investigates the robustness of modules with different design options using finite element simulations. The simulation results serve as input for a lifetime prediction model. Key strategies to enhance module robustness are identified, including the use of sensors with increased thicknesses, an increased number of solder bumps, and optimized adhesive patterns. Experimental tests are later conducted to validate the simulation findings.


\section{Thermomechanical Study of HGTD Module} \label{simulation}


Due to the impracticality of testing the robustness of many modules with varying designs before full-scale production for the HGTD, finite element analysis was employed. This approach required the development of a realistic module model and the simulation of thermal cycling conditions. The resulting stress profiles within the solder joints and the subsequent lifetime predictions informed the development of new design solutions, which were later implemented and experimentally validated.
    
\subsection{ Finite Element Model of HGTD Module Structure and Loading of Thermal Cycle Conditions} \label{sec:models}
    
\begin{figure}[h]
        \centering
        \includegraphics[width=0.5\linewidth]{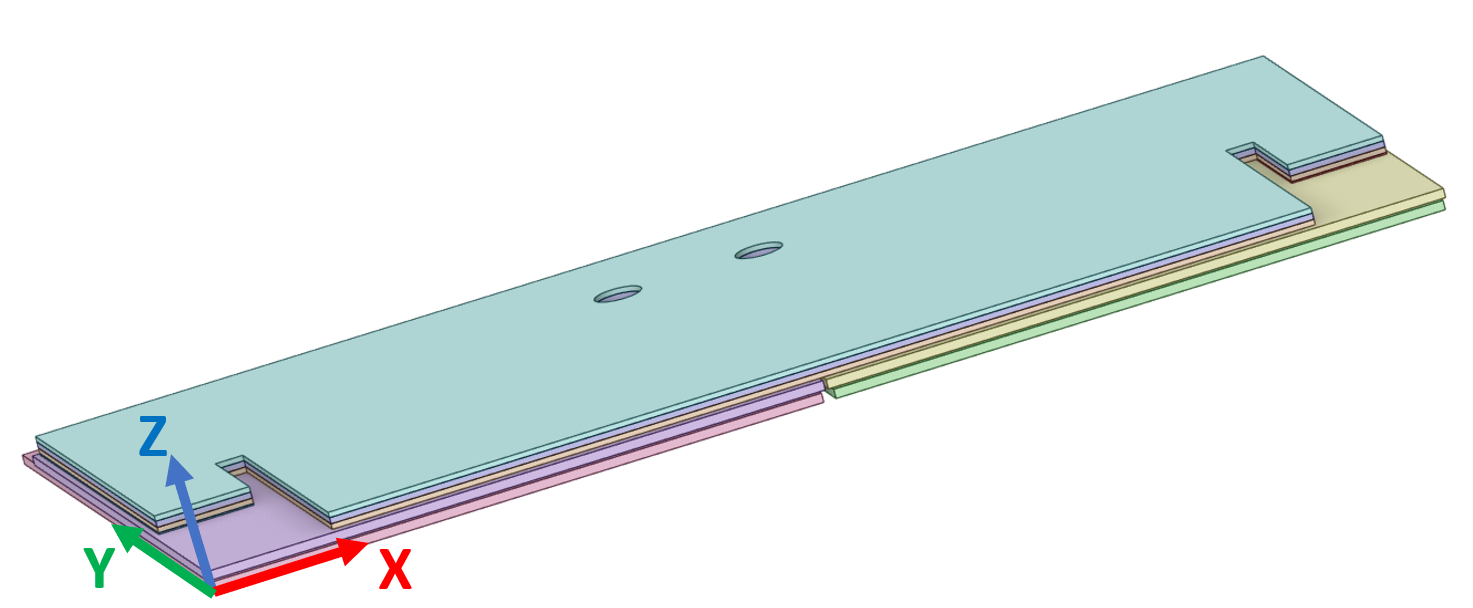}
        \caption{A 3-D model of the HGTD module was developed for simulation purposes. The assembly, from top to bottom, consists of a multi-layer PCB, an adhesive layer, two sensors, an array of bumps, and two ASICs. The PCB structure employs a simplified layer representation and an approximated geometric shape to help efficient simulation. The coordinate system is also defined.}
        \label{fig:module_baseline}
    \end{figure}

A simplified 3-D model of an HGTD module was developed for FEM simulation based on the design specifications, as illustrated in Figure \ref{fig:module_baseline} and summarized in Table \ref{t1}. The PCB is represented as a three-layer structure, with a middle layer of copper alloy (simplified from the actual six copper layers) serving as the circuit layer, and top and bottom layers of Kapton (polyimide, PI) providing mechanical support and protection.

    \begin{table}[h]
    \centering
    \caption{\label{t1}Layer layout of the 3-D model of HGTD module for simulation. Three simplified layers are employed for the PCB.}  
    \begin{tabular*}{0.8\textwidth}{@{\extracolsep{\fill}}l c c}
    \toprule    
    \textbf{Layer} & \textbf{Material} & \textbf{Thickness ($\mu$m)}\\  
    \midrule 
    PCB layer-1 & Polyimide (PI) & 175\\
    PCB layer-2 & Copper alloy & 200\\
    PCB layer-3 & Polyimide (PI) & 175\\
 
    Glue & Epoxy resin (Araldite 2011) & 50\\

    Sensor & Silicon & 300\\

    Bump & $\mathrm{Sn_{3.5}Ag}$ & 50\\

    ASIC & Silicon & 300\\
    \bottomrule 
    \end{tabular*}  
    
    \end{table} 

    \begin{figure}[h]
        \centering
        \subfloat[\label{microscope}]{\includegraphics[width = 0.3\textwidth]{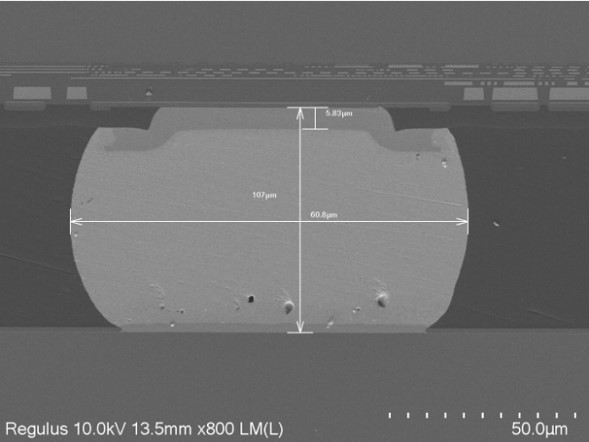}}
        \hfill
        \subfloat[\label{model}]{\includegraphics[width = 0.3\textwidth]{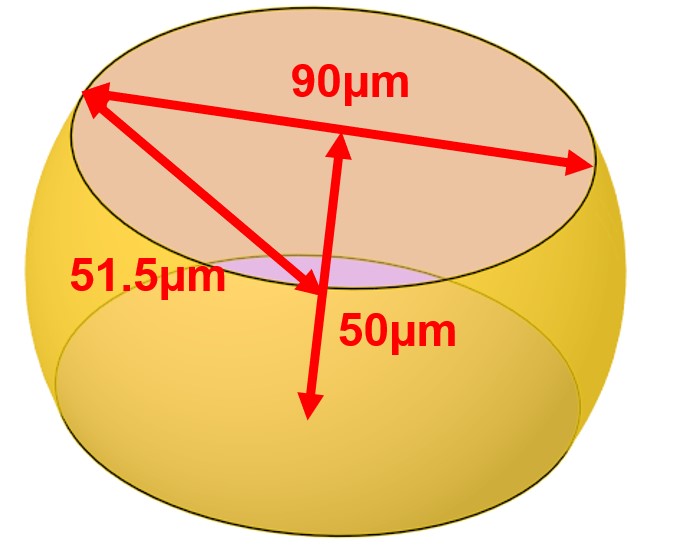}}
        \hfill
        \subfloat[\label{bumparray}]{\includegraphics[width = 0.3\textwidth]{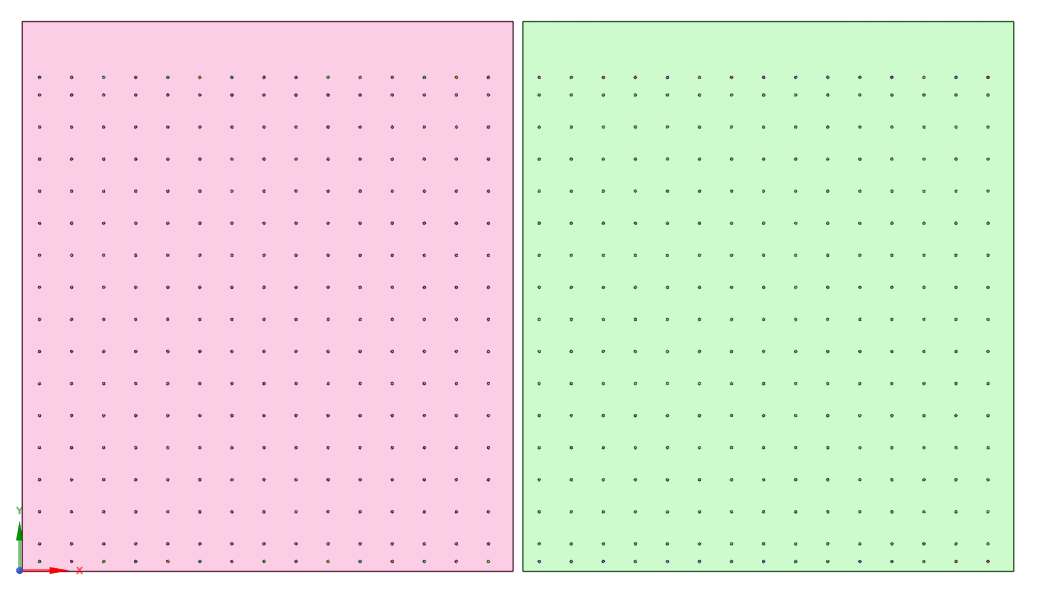}}
        \hfill
    \caption{The bump structure in HGTD modules. (a) A bump cross section observed in electron microscope with a measurement of some dimensions. (b) A single bump in the 3-D model, which is a sphere of radius 51.5~$\mathrm{{\mu}m}$ and cut by top and bottom part symmetrically to keep 50~$\mathrm{{\mu}m}$ height. (c) Bump array shown relative to ASICs in the 3-D model. Besides 15$\times$15 bumps on all pixels in the middle of the hybrid, there are 15 extra bumps on the top and bottom edge each. These bumps are not part of the readout circuit, instead they are used for protection of bumps on pixels. The distance between two adjacent bumps is 1.3~mm.}
    \label{bump}
    \end{figure}    
    
    The bump structure is created according to a cross-section photo in electron microscope and module specification shown in Figure~\ref{bump}. The bump material used in the HGTD module is $\mathrm{Sn_{3.5}Ag}$. Under thermal stress, the inelastic deformation of the solder joints mainly includes plastic deformation and creep deformation. From the perspective of nonlinear continuum mechanics, both plastic deformation and creep strain are related to dislocation movement. Hence, viscoplasticity is often used to describe the nonlinear characteristics of solder and characterize its mechanical behaviors~\cite{viscomodel}. 
    The Anand model is a widely used viscoplastic model~\cite{visco,anandmodel} that is provided in the Ansys Software~\cite{ansys} used in this paper. 
    The Anand model can be considered as a combination of a flow Equation~\ref{flow_equation}, a strain hardening evolution Equation~\ref{evolution_equation} and a saturation value Equation~\ref{definition_s}:
    \begin{equation}\label{flow_equation}
        \frac{d\epsilon_p}{dt}=Ae^{-\frac{Q}{RT}}\left[\sinh\left(\xi\frac{\sigma}{S}\right)\right]^{\frac{1}{m}},
    \end{equation}\begin{equation}\label{evolution_equation}
        \frac{dS}{dt}=\left\{h_0\left|1-\frac{S}{S^*} \right|^{a}\mathrm{sign}\left(1-\frac{S}{S^*}\right)\right\}\frac{d\epsilon_p}{dt},
    \end{equation}\begin{equation}\label{definition_s}
        S^*=\hat{S}\left[\frac{\frac{d\epsilon_p}{dt}e^{-\frac{Q}{RT}}}{A}\right]^n.
    \end{equation}
    The three equations together solve for three variables: plastic strain ($\epsilon_p$), stress ($\sigma$), and deformation resistance ($S$) as functions of time. 

  The specific Anand model parameters for the $\mathrm{Sn_{3.5}Ag}$ solder used in this study and their definitions are listed in Table \ref{t2}~\cite{anand}, while the general material properties for all components are provided in Table \ref{t4}. The constitutive behavior for the remaining materials is described by linear elastic models. All materials, except silicon, are treated as isotropic.

     \begin{table}[!h]
    \centering 
    \caption{\label{t2}Anand model parameters for $\mathrm{Sn_{3.5}Ag}$.}
    \begin{tabular*}{0.8\textwidth}{@{\extracolsep{\fill}}l c}
    \toprule 
    \textbf{Model parameter} & \textbf{Value}\\
    \midrule
    Initial deformation resistance $S_0$ & 2.3165~MPa\\

    Activation energy than universal gas constant $Q/R$ &  $10279~\SI{}{\degreeCelsius}$\\

    Pre-exponential factor $A$ & 1.7702$\times 10^{5}$$~\mathrm{s^{-1}}$\\

    Multiplier of stress $\xi$ & 7\\

    Strain rate sensitivity of stress $m$ & 0.207\\

    Hardening/Softening constant $h_{0}$ & 27782~MPa\\

    Coefficient for deformation resistance saturation $\hat{S}$ & 52.4~MPa\\

    Strain rate sensitivity of saturation $n$ & 0.0177\\

    Strain rate sensitivity of hardening or softening $a$ & 1.6\\
    \bottomrule
    \end{tabular*}
    
    \end{table}

    \begin{table}[!h]
    \centering
     
    \caption{\label{t4}The material properties employed in the simulation account for the crystalline nature of silicon, which is modeled as an anisotropic material. All other materials are treated as isotropic.}
    \begin{tabular*}{0.8\textwidth}{@{\extracolsep{\fill}}l c c c}
    \toprule
    \textbf{Material} &  \textbf{CTE} & \textbf{ Young's} & \textbf{ Poisson's}  \\      
    & \textbf{(ppm/\SI{}{\degreeCelsius} )} & \textbf{modulus (GPa)} & \textbf{ratio} \\  
    \midrule
    Copper &  18 & 110 & 0.34 \\
  
    Epoxy resin & 70 & 3.78 & 0.35 \\

    Polyimide & 20 & 2.8 & 0.34 \\

    $\mathrm{Sn_{3.5}Ag}$ & 19 & Anand model & 0.4 \\

     & & 169(Ex) & 0.064($\mathrm{\nu_{xy}}$) \\
    Silicon & 2.578 & 169(Ey) & 0.36($\mathrm{\nu_{yz}}$) \\
     & & 130(Ez) & 0.36($\mathrm{\nu_{zx}}$) \\
    \bottomrule
    \end{tabular*}
    
    \label{tab:material}
    \end{table}   

\begin{figure}[!h]
        \centering
        \subfloat[\label{gluepattern_real}]{\includegraphics[width = 0.45\textwidth]{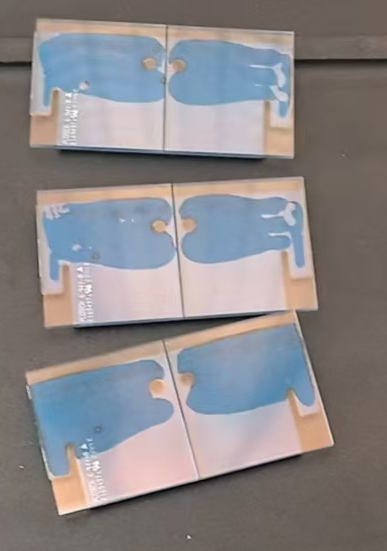}}
        \hfill
        \subfloat[\label{gluepattern}]{\includegraphics[width = 0.45\textwidth]{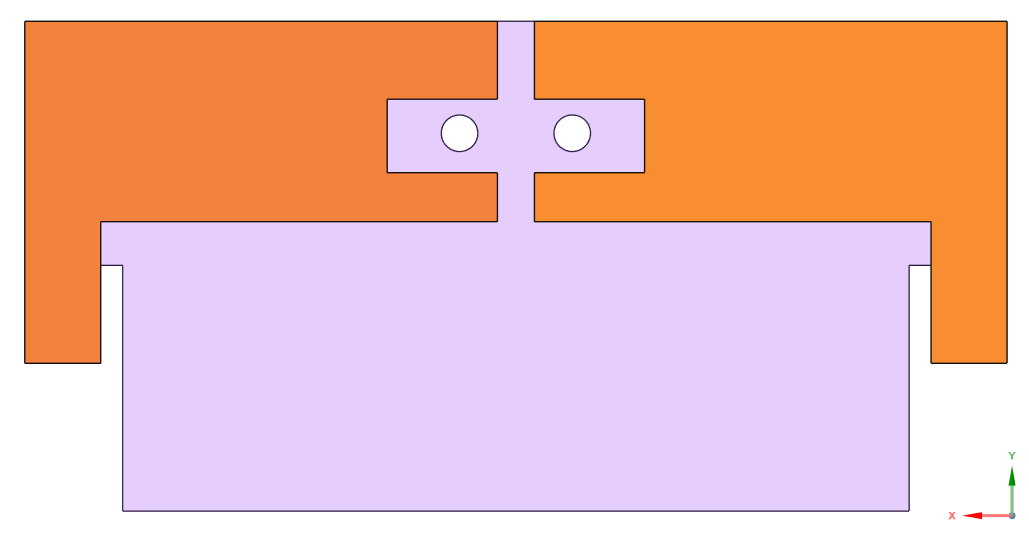}}
        \hfill
    \caption{Model for glue pattern in the HGTD module. (a) Dummy modules assembled on glass, showing the glue pattern in blue. (b) Glue pattern (in orange) in the 3-D model relative to PCB position (in mauve). In reality the glue does not cover all the glue area in the model.}
    \label{glue}
    \end{figure}

The adhesive pattern and its corresponding simulation model are depicted in Figure~\ref{glue}. The pattern provides coverage across the module's upper section.
The top edge of the PCB must maintain sufficient flatness and stability to furnish a robust support structure for wire bonding. This requirement also applies to the two wing sections of the PCB, which are designed to be mounted onto the detector. Guaranteeing adhesive coverage in these critical areas is essential.

Owing to the geometric asymmetry of the module, the full 3-D model was utilized in the simulation. A refined mesh was applied to the bumps, with each bump discretized into 80–90 elements to ensure computational accuracy. The entire model contains approximately 350,000 elements and 880,000 nodes. All inter-layer contacts are defined as bonded contacts. A fixed constraint was applied to one corner vertex of the assembly, while no other constraints were imposed, allowing the module to deform freely under simulated conditions.

\begin{figure}[!h]
        \centering
        \includegraphics[width=0.8\linewidth]{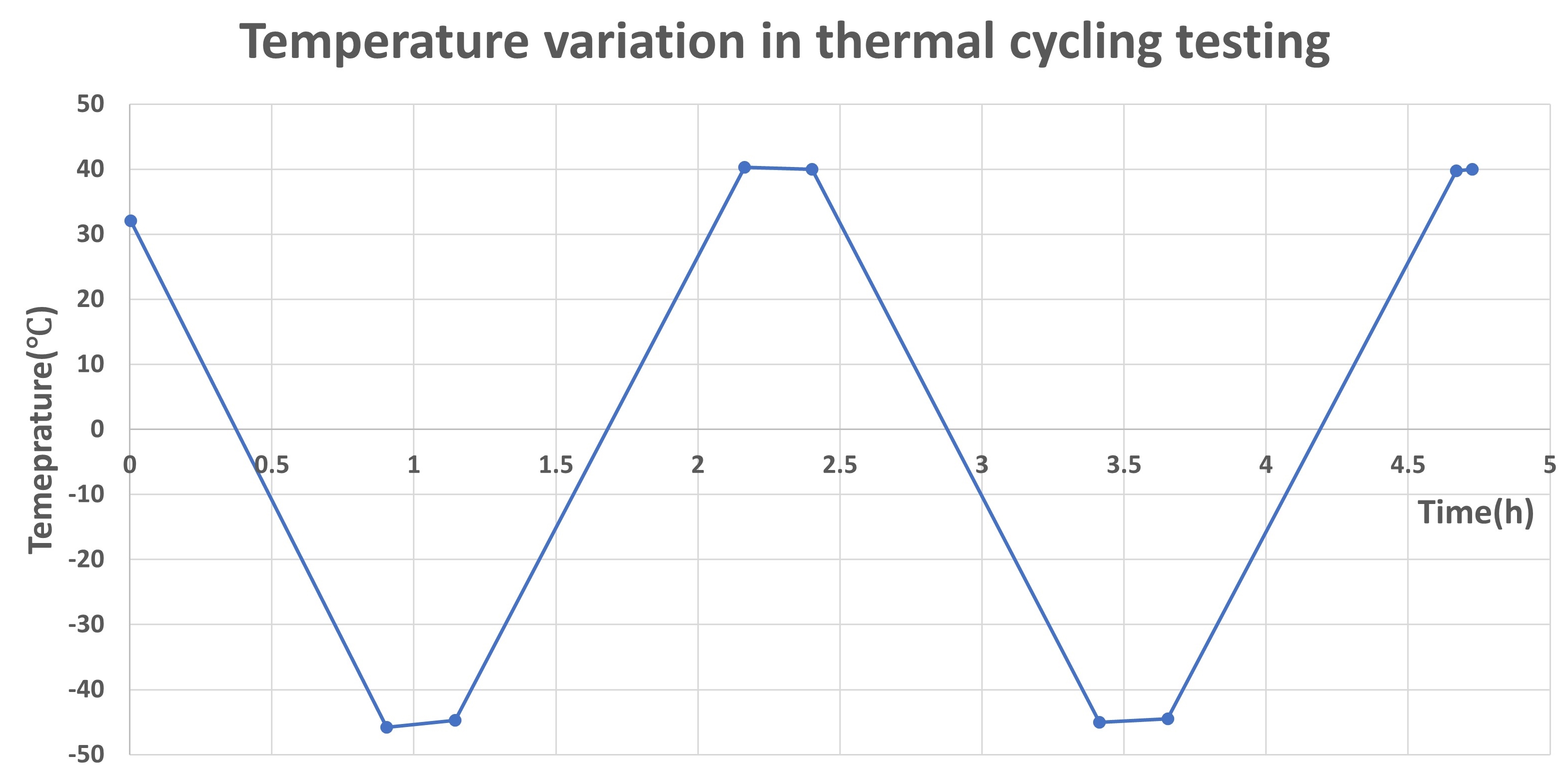}
        \caption{Temperature variation detected in thermal cycling testing, ranging from -45~$^{\circ}\mathrm{C}$ to 40~$^{\circ}\mathrm{C}$. The temperature is maintained for about 15 minutes at the lowest and highest temperature in each cycle.}
        \label{fig:tc_real}
    \end{figure}

\begin{figure}[!h]
         \centering
         \includegraphics[width=0.8\linewidth]{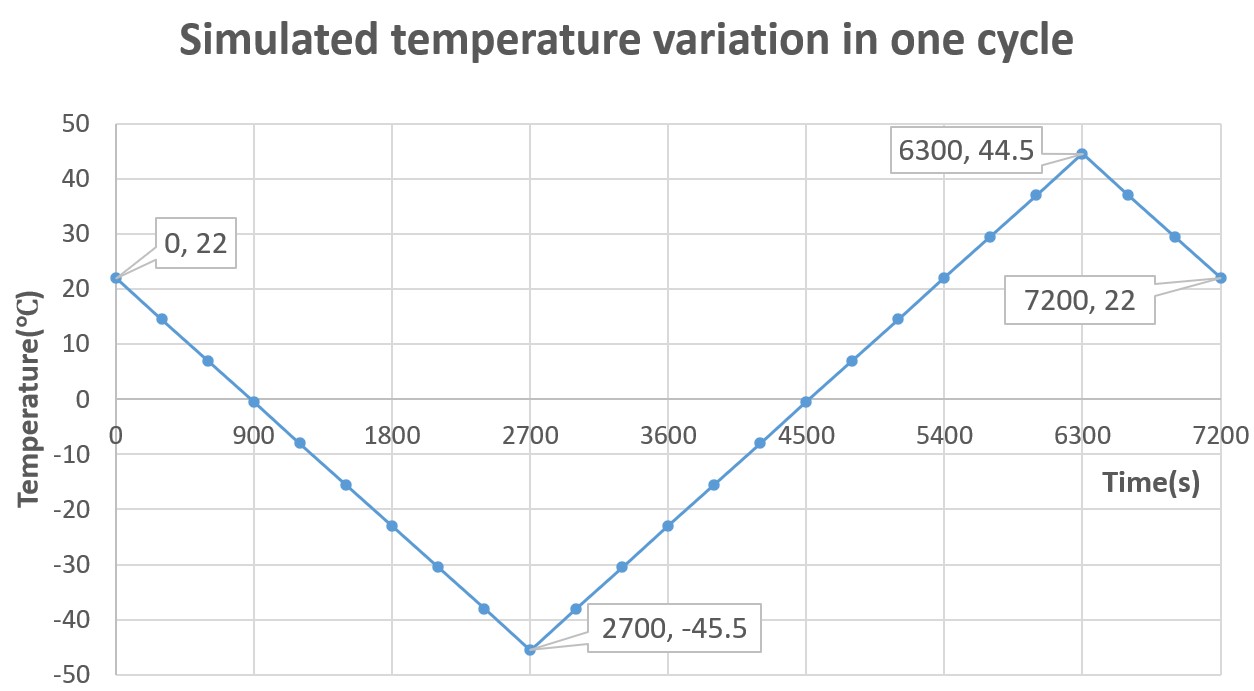}
         \caption{Temperature condition applied in simulation. Each dot represents a time step of 300 seconds, during which the absolute temperature change is 7.5~$^{\circ}\mathrm{C}$.}
         \label{temperature}
     \end{figure}

The thermal cycling profile, illustrated in Figure \ref{fig:tc_real}, is applied at a slow rate: each cycle spans from a minimum of -45~$^{\circ}\mathrm{C}$ to a maximum of 40~$^{\circ}\mathrm{C}$ over a period of 2.5 hours. This temperature range extends beyond the nominal operating conditions—which range from -30~$^{\circ}\mathrm{C}$ to room temperature (22~$^{\circ}\mathrm{C}$)—in order to incorporate a safety margin and ensure robust module performance under accelerated stress.
     The thermal cycling conditions applied in the simulation are shown in Figure~\ref{temperature}. The temperature range is -45.5~$^{\circ}\mathrm{C}$ to 44.5~$^{\circ}\mathrm{C}$ (a little larger range than in real testing), with no peak holding time and uniform temperature change in two hours. 24 load steps are used and each step lasts 300~s with a temperature change of 7.5~$^{\circ}\mathrm{C}$. Since the rate of temperature change is low enough, the temperature load is applied to all units directly. The default value of the strain reference temperature is 22~$^{\circ}\mathrm{C}$.
    
\subsection{Lifetime Prediction Model for Solder Joint Structure}

The Coffin--Manson model, based on mechanical strain amplitude, is widely employed for the fatigue life prediction of solder joints~\cite{coffin-manson_use1,coffin-manson_use2}. This empirical relationship establishes a correlation between cyclic durability and strain amplitude for specific materials and temperature conditions, expressed as:
\begin{equation}
    \label{equ:Coffin-Manson}
    N_{f} = \frac{1}{2} \left( \frac{\Delta\epsilon}{2\epsilon_f'} \right)^{\frac{1}{c}}.
\end{equation}
In Equation~\ref{equ:Coffin-Manson}, $N_{f}$ represents the mean number of cycles to failure (corresponding to a 50\% probability of failure for any bump), $\Delta\epsilon/2$ is the cyclic mechanical strain amplitude, $\epsilon_f' = 0.325$ is the fatigue ductility coefficient for the solder joint material, and $c$ is the fatigue ductility exponent.

Several modified versions of Equation~\ref{equ:Coffin-Manson} were developed to account for different temperature ranges and cycling frequencies. One successful formulation, which demonstrates that the fatigue ductility exponent $c$ is the primary governing parameter, is given by Equation~\ref{equ:c}~\cite{coffin-manson1,coffin-manson2}. This model incorporates a linear temperature dependence and a logarithmic frequency dependence:
\begin{equation}
    \label{equ:c}
    c = -0.442 - 6 \times 10^{-4} T_s + 1.74 \times 10^{-2} \ln(1 + f), \quad 1 \leq f \leq 1000.
\end{equation}
Here, $T_s$ denotes the mean cyclic solder temperature in $^{\circ}\mathrm{C}$, and $f$ is the cycling frequency expressed in cycles per day.

In this study, the thermal cycling profile ranges from \SI{-45.5}{\degreeCelsius} to \SI{44.5}{\degreeCelsius}, resulting in a mean solder temperature $T_s = \SI{-0.5}{\degreeCelsius}$. Given this negligible temperature offset, the value of $c$ depends primarily on the cycling frequency. With a cycle period of two hours ($f = 12$ cycles per day), the calculated exponent is $c = -0.397$. This yields a relationship where larger strain amplitudes correspond to shorter mean cycles to failure. The lifetime of the module/hybrid is governed by the minimum fatigue lifetime among all its bumps, as determined by the Coffin-Manson model. This critical bump, exhibiting the shortest lifetime, is consequently prognosticated to be the primary failure initiation site. The total mechanical strain range per cycle ($\Delta\epsilon$) was evaluated through FEM simulation using Ansys. The specific value was extracted from the simulation results by employing the \texttt{EPTOINT} command to obtain the strain at the element integration points.

    \subsection{Simulation Results and Proposed HGTD Module Designs}
    \label{sec:setup}

\begin{figure}[h]
        \centering
        \subfloat[]{\includegraphics[width = 0.55\textwidth]{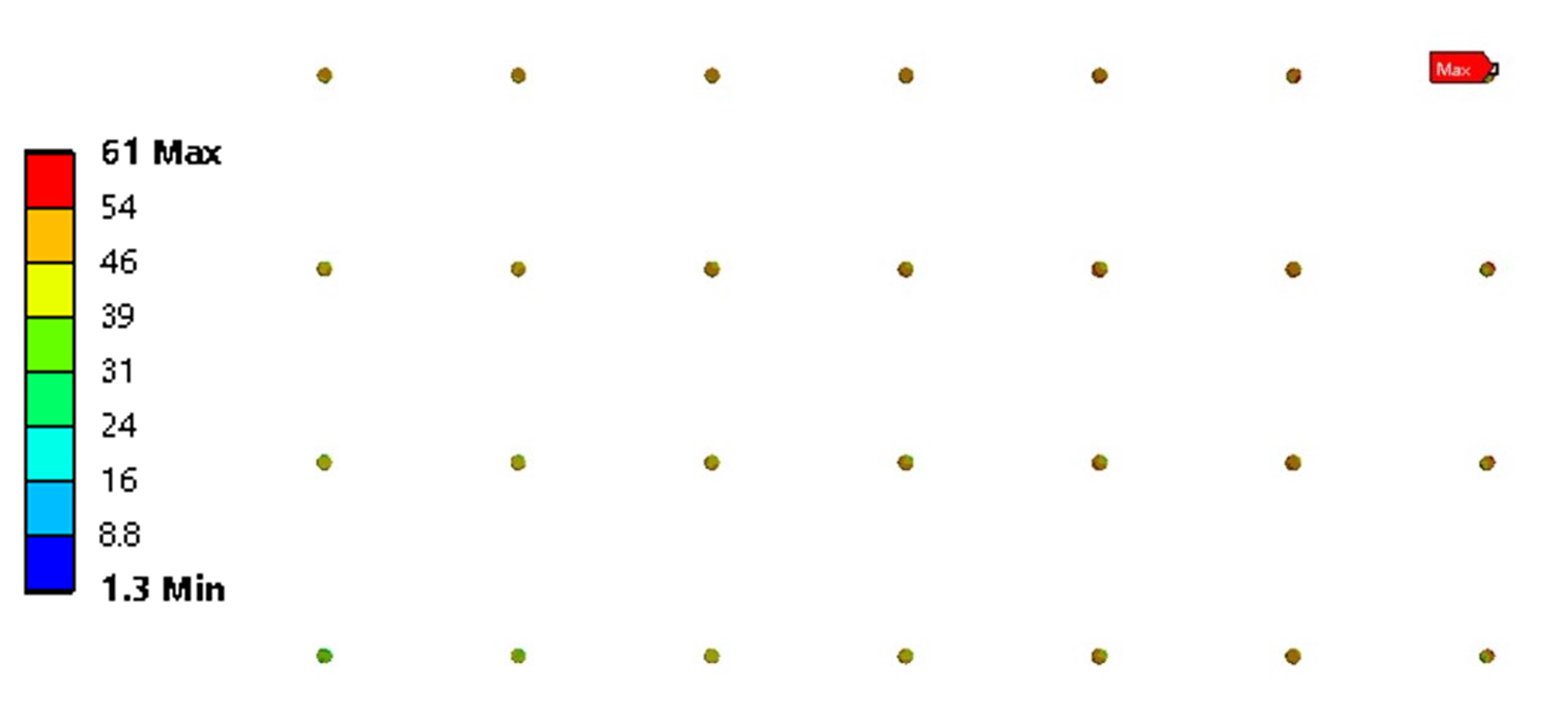}}
        \hfill
        \subfloat[]{\includegraphics[width = 0.4\textwidth]{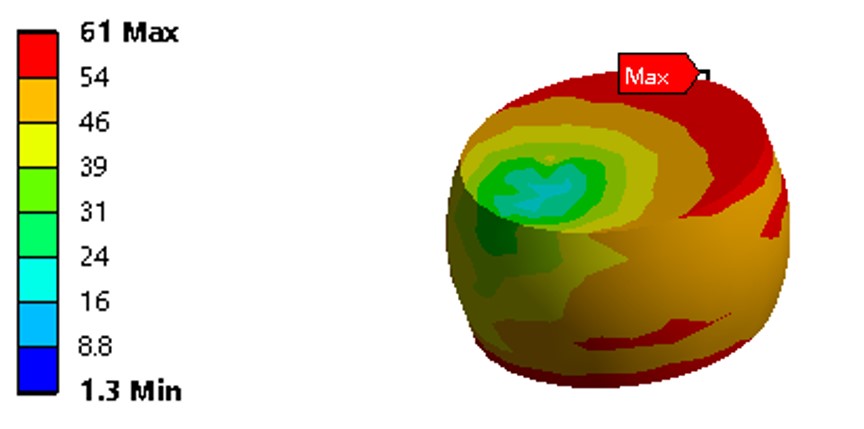}}
        \hfill
        \caption{Bump von-Mises stress (unit: MPa) distribution at lowest temperature (greatest thermal strain). (a) Bump Von-Mises stress distribution on the top right corner part of the module. Maximum stress appears at the corner bump while bumps to the center are under smaller stress. (b) Cloud diagram of Von-Mises stress in a single bump with maximum stress. Contact surfaces to sensor and ASIC are under greater stress and the maximum appears at the surface contacting the sensor.}
        \label{bump stress}
    \end{figure}

Figure~\ref{bump stress} presents the von-Mises stress distribution in the solder bumps, derived from the finite element model of the A30 hybrid module (specifications provided in Table~\ref{tab:gr_full}). Analysis of the full model indicates that although the two hybrids are nearly symmetric, the stress is pronouncedly asymmetry between the top and bottom regions of each hybrid. Bumps located in the bottom region experience significantly lower stress levels than those in the top region.
Within the top region, the stress magnitude increases progressively with distance from the center, culminating in maximum values at the corner bumps. This distribution arises from the thermal cycling conditions: due to their higher CTE, materials such as copper and PI cause the PCB to undergo a thermal strain that is proportional to the distance from its geometric center. This deformation is transmitted through the adhesive layer to the sensor, thereby inducing shear stress in the solder bumps. The bottom region remains largely unaffected due to the absence of adhesive.
Consequently, the corner bumps in the top region are subjected to the highest mechanical strain, particularly at their top and bottom contact interfaces. According to lifetime prediction models, these locations are the most susceptible to early failure, which provides a mechanical explanation for the observed initial bump connection failures predominantly occurring at these sites.
    
\begin{figure}[t]
        \centering
        \includegraphics[width=0.5\linewidth]{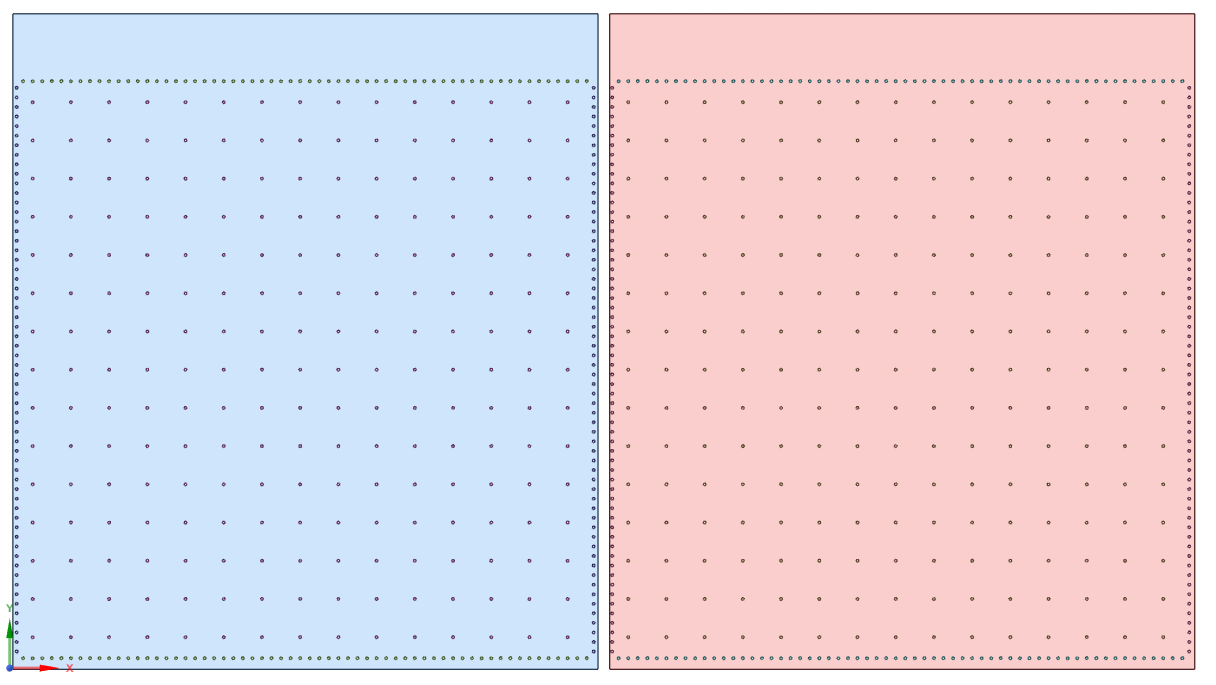}
        \caption{Bump array in the proposed design shown relative to ASICs in the 3-D model. In this design, there are 60 bumps on each of the four edges on the guard ring, leading to a total number of 240 guard ring bumps on each hybrid. The distance between adjacent bumps on the guard ring is 0.325~mm.}
        \label{fig:more bumps}
    \end{figure}

Based on the analysis of stress distributions, several new module designs are proposed to mitigate bump stress, primarily involving modifications to the sensor, bump configuration, and adhesive application.
The primary source of deformation is the PCB\@. Consequently, employing a thicker sensor is hypothesized to reduce the strain transfer from the PCB\@. A sensor thickness of \SI{775}{\micro\meter} (the standard full-wafer thickness) is proposed to replace the baseline thinned sensor of \SI{300}{\micro\meter}; this change is not expected to adversely affect detector performance.
Furthermore, increasing the number of bumps between the sensor and the ASIC can help distribute the mechanical stress concentrated on the pixel bumps, thereby reducing the stress per bump. However, adding bumps directly to the readout pads would increase parasitic capacitance and impair electrical performance. The proposed solution is to add supplemental bumps exclusively on the guard ring of the hybrid (Figure~\ref{fig:more bumps}). This design increases the total number of guard ring bumps on a single hybrid from $15\times2 = 30$ to $60\times4 = 240$, significantly enhancing the bump density, particularly in the corner regions.

Simulation results presented in Table~\ref{tab:gr_full} validate both design strategies, demonstrating a substantial increase in predicted lifetime for the new configurations with thicker sensors and more guard ring bumps. The hybrid types A and B mark the difference in sensor thickness (A for \SI{300}{\micro\meter} and B for \SI{775}{\micro\meter}), while the number after the Letter represents the total guard ring bumps. The predicted lifetime of 16 cycles for the A30 module configuration is consistent with thermal cycling test results, where multiple modules exhibited bump connection failures after only a few cycles. This agreement between simulation and experiment confirms the accuracy of the finite element model.    

\begin{table}[t]
    \centering
    \caption{Impact of sensor thickness and number of guard ring bumps on module lifetime. Configurations with a thicker sensor (775~$\mu$m) and a greater number of guard ring bumps show a significantly prolonged lifetime.}
    \begin{tabular*}{0.9\textwidth}{@{\extracolsep{\fill}}l c c c c}
        \toprule
        \textbf{Hybrid} & \multicolumn{2}{c}{\textbf{Type A}} & \multicolumn{2}{c}{\textbf{Type B}} \\
        \cmidrule(lr){2-3} \cmidrule(lr){4-5}
        & \textbf{A30} & \textbf{A240} & \textbf{B30} & \textbf{B240} \\
        \midrule
        Sensor thickness & \multicolumn{2}{c}{300~$\mu$m} & \multicolumn{2}{c}{775~$\mu$m} \\
        Guard ring bumps & 30 & 240 & 30 & 240 \\
        Total bumps & 255 & 465 & 255 & 465 \\
        \midrule
        Lifetime (cycles) & 16 & 108 & 119 & 393 \\
        \bottomrule
    \end{tabular*}
    
    \label{tab:gr_full}
\end{table}

Modifying the adhesive pattern presents a further strategy to mitigate deformation transfer from the PCB, as the adhesive serves as the mechanical interface between the sensors and the PCB\@. The prior observation that bumps in unglued regions experience lower stress suggests that reducing the adhesive coverage area could be beneficial. To maintain structural stability, enough adhesive must be retained to secure the top edge (for wire bonding) and the wing sections (for integration into the detector).

\begin{figure}[!h]
        \centering
        \subfloat[]{\includegraphics[width = 0.45\textwidth]{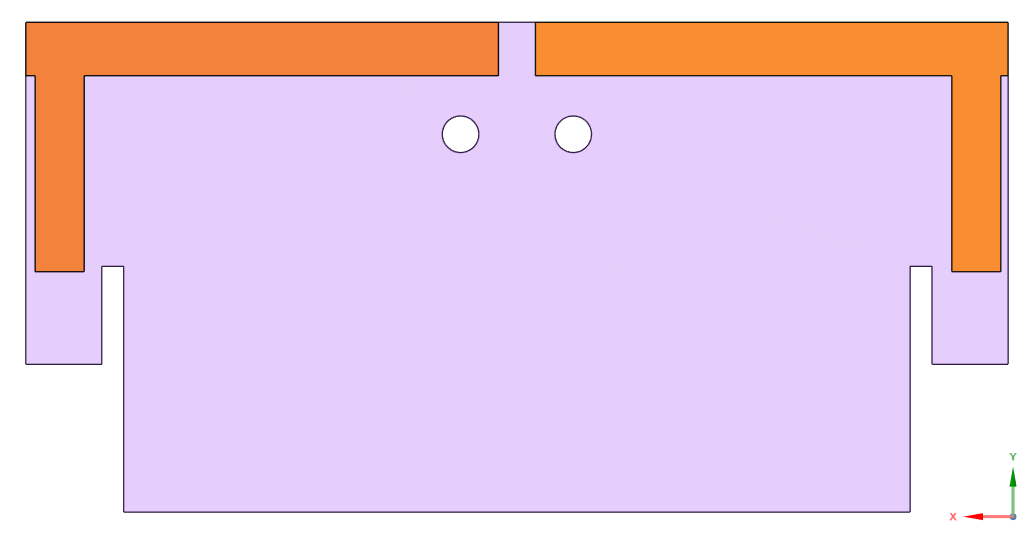}}
        \hfill
        \subfloat[]{\includegraphics[width = 0.45\textwidth]{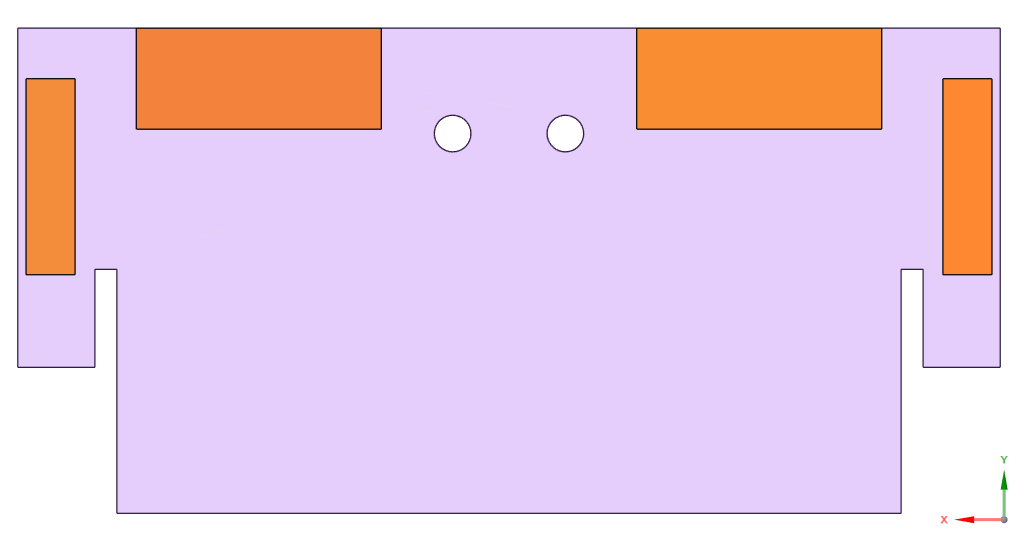}}
        \hfill
        \caption{Models for glue patterns (in orange) where glue covers areas for supporting wire-bonding and fixture into  the detector. (a) Pattern 1. (b) Pattern 2. The glue regions are split apart in pattern B while the total amount of glue is similar in both patterns.}
        \label{proposal glue}
    \end{figure}

\begin{figure}[h]
        \centering
        \subfloat[\label{stress_sj}]{\includegraphics[width = 0.9\textwidth]{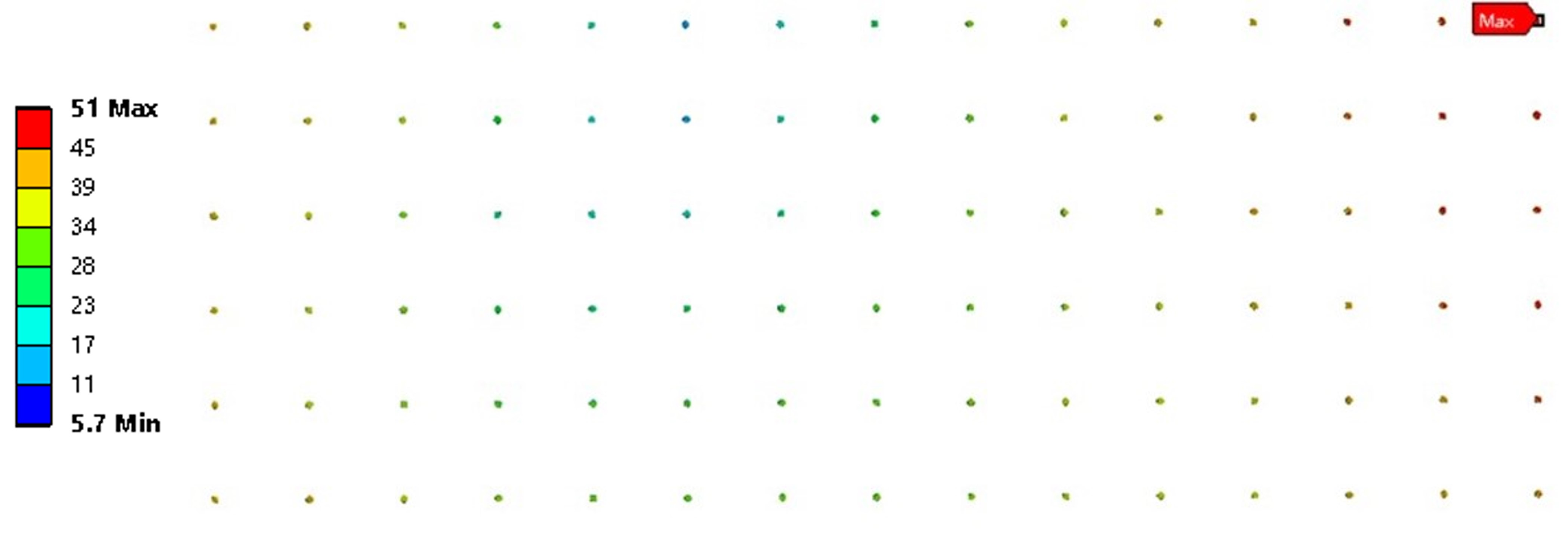}}
        \hfill
        \subfloat[\label{stress_j}]{\includegraphics[width = 0.9\textwidth]{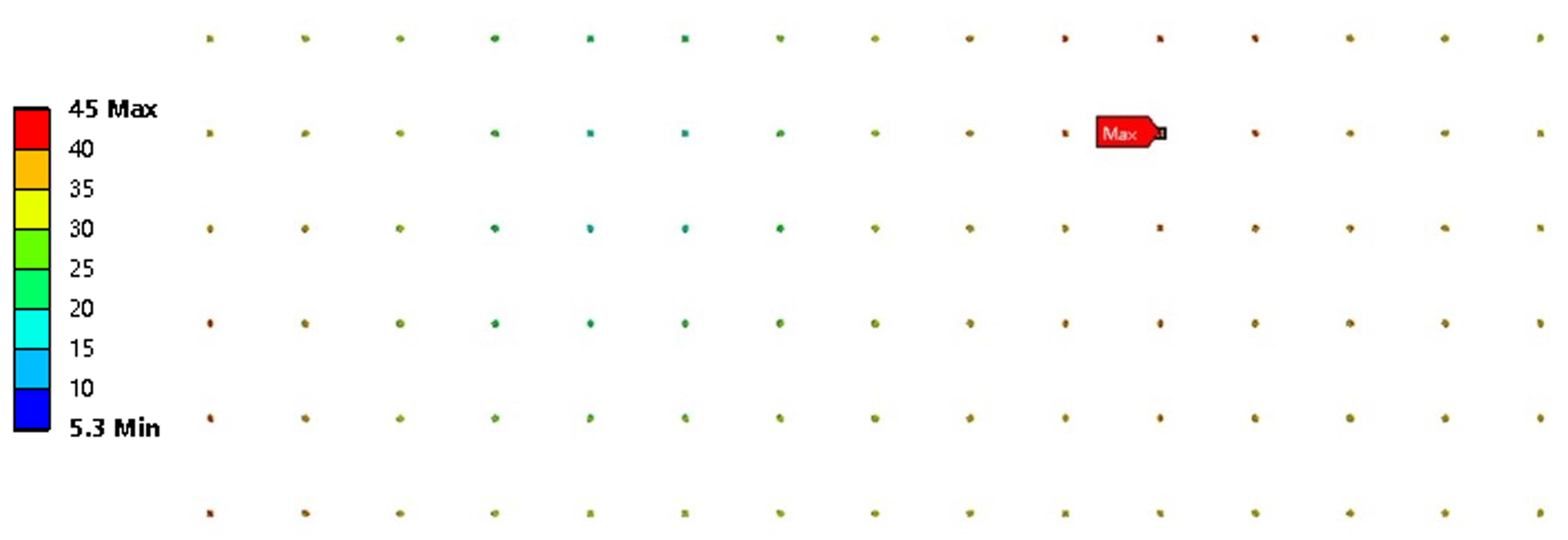}}
        \hfill
        \caption{Equivalent stress (unit:MPa) distribution in two proposed glue patterns 1 and 2. Top part of right hybrid is shown in both cases. (a) Pattern 1. The maximum stress still appears at the corner bump. (b) Pattern 2. The maximum stress does not appear at the corner any more as there is no glue there.}
        \label{stress distribution}
    \end{figure}

Two tentative adhesive patterns, shown in Figure~\ref{proposal glue}, were simulated; however, their feasibility for mass production requires further validation. The simulation results, obtained using the new B240 hybrid design, are presented in Table~\ref{tab:gluepattern}. A simulation with no adhesive, which completely avoids deformation transfer from the PCB, shows a dramatic increase in module lifetime. This result strongly supports the hypothesis that bump joint failures in HGTD modules during thermal cycling are primarily driven by PCB deformation.
Although the total adhesive mass is similar in both proposed patterns, the resulting module lifetime differs significantly. Both Pattern 1 and Pattern 2 yield a longer lifetime than the baseline configuration, with Pattern 2 exhibiting a particularly substantial improvement. Analysis of the von Mises stress distribution in the bumps (Figure~\ref{stress_j}) reveals that in Pattern 2, the maximum stress is redistributed away from the hybrid corners. The revised adhesive pattern shifts a greater proportion of the stress to bumps located beneath the adhesive, where the PCB's deformation is minimized due to proximity to its geometric center.
In contrast, while Pattern 1 also reduces adhesive usage, it retains adhesive over the corner bumps transmitting the PCB strain. As these corners remain the point of highest stress concentration (Figure~\ref{stress_sj}), the improvement in lifetime, though present, is less pronounced than that achieved by Pattern 2.

    \begin{table}[!h]
    \centering
    \caption{Comparison of different glue patterns based on B240 hybrid configurations. Increased lifetime is expected with the two proposed glue patterns.}
   \begin{tabular*}{0.95\textwidth}{@{\extracolsep{\fill}}lcccc}
   \toprule
        & \textbf{No glue (hybrid)}  & \textbf{Baseline pattern} & \textbf{Pattern 1} & \textbf{Pattern 2}\\
        \midrule
        \text{Lifetime (cycles)} & \text{1240749} & \text{393} & \text{566} & \text{3029} \\
        \text{Glue mass per module (mg)} & \text{0} & \text{18.8} & \text{6.7} & \text{6.6}\\
        \bottomrule
    \end{tabular*}

        \label{tab:gluepattern}
\end{table}

\section{Experimental Validation}
Selected design configurations from the simulation study were implemented in prototype modules and subjected to thermal cycling tests to validate the predictive power of the finite element model. Two module versions were tested. The first Type A30 featured a thin \SI{300}{\micro\meter} sensor and 30 guard ring bumps per hybrid, leading to a simulated mean lifetime of 16 cycles. The second Type B30 incorporated a thick \SI{775}{\micro\meter} sensor while retaining the 30-bump configuration, resulting in a simulated mean lifetime of 119 cycles.

Thermal cycling was performed in temperature-controlled environmental chambers. Following increments of several cycles, modules were extracted and evaluated for bump connectivity at room temperature. A combined testing approach, employing both radioactive source measurements and a threshold voltage analysis, was used to generate bump connectivity maps. The source measurement method applies a high voltage to the sensor and measures the electrical response to particles from the radioactive source to determine the connection status. The threshold voltage method involves a threshold voltage scan for different injected charges of 36 DAC (12~fC) and 12 DAC (4~fC). By calculating the difference between two threshold voltage values, disconnected bump can be identified (small difference value). To enhance statistical significance given the limited number of sample modules, each module was treated as two independent hybrids for testing and analysis.

The HGTD operational lifetime requirement, accounting for detector commissioning, module replacements, and nominal data-taking periods, has been established as equivalent to 36 thermal cycles. This value serves as the qualification standard for module testing. Initial testing of Type A30 modules exhibited premature failure, with numerous bump disconnections observed within the first 10--20 cycles. In response, Type B30 modules were fabricated using \SI{775}{\micro\meter} thick sensors while maintaining the original bump configuration. This enhanced design demonstrated significantly improved reliability, with no bump connection failures observed before 30 cycles across a sample of 46 hybrids. As detailed in Table~\ref{thick_tc2}, bump disconnections were observed in 14 Type B30 hybrids during testing. The remaining 32 hybrids, documented in Table~\ref{thick_tc1}, exhibited no disconnections through the final completed cycles. These results are consistent with the simulated prediction of a 119-cycle mean lifetime, substantially exceeding both the HGTD requirement and the performance of Type A30.

In both Type A30 and B30 modules, most of bump failures occurred at the corners, aligning with the stress concentration regions identified in simulations. Across all tested modules, only a minimal fraction of bumps (20, representing 0.2\% of the total) exhibited disconnections--18 of which occurred at cycle counts significantly beyond the 36-cycle requirement. Projecting this failure rate to the full HGTD system comprising 8032 modules, fewer than 1000 bumps are anticipated to disconnect by the end of the operational lifetime.

    \begin{table}[!h]
        \centering
        \caption{Bump connection failures on the full set of 46 Type B30 hybrids. Failed hybrids are hybrids with any disconnected bump(s). Number of disconnected bumps represents the new disconnected bumps at a certain cycle count. }
        \begin{tabular*}{0.9\textwidth}{@{\extracolsep{\fill}}lcccccccc}
        \toprule
              \textbf{Tested cycle number} & \textbf{30} & \textbf{60} & \textbf{75} & \textbf{90} & \textbf{105} & \textbf{120} & \textbf{180} & \textbf{Total} \\
          \midrule
          Number of failed hybrids & 0 & 2 & 3 & 3 & 0 & 4 & 2 & 14 \\
          Number of disconnected bumps & 0 & 2 & 5 & 5 & 0 & 6 & 2 & 20\\
          \bottomrule
        \end{tabular*}
        
      \label{thick_tc2}
    \end{table}
    \begin{table}[!h]
        \centering
        \caption{Number of hybrids without bump detachment as a function of the total number of thermal cycles completed, e.g., five hybrids survived 150 thermal cycles without disconnects.}
        \begin{tabular*}{0.7\textwidth}{@{\extracolsep{\fill}}lccccc}
        \toprule
          \textbf{Completed cycle number} & \textbf{75} & \textbf{120} & \textbf{150} & \textbf{180} & \textbf{Total} \\
          \midrule
          Number of hybrids & 4 & 14 & 5 & 9 & 32 \\
          \bottomrule
        \end{tabular*}
        
      \label{thick_tc1}
    \end{table}

\begin{figure}[!h]
        \centering
        \includegraphics[width=0.5\linewidth]{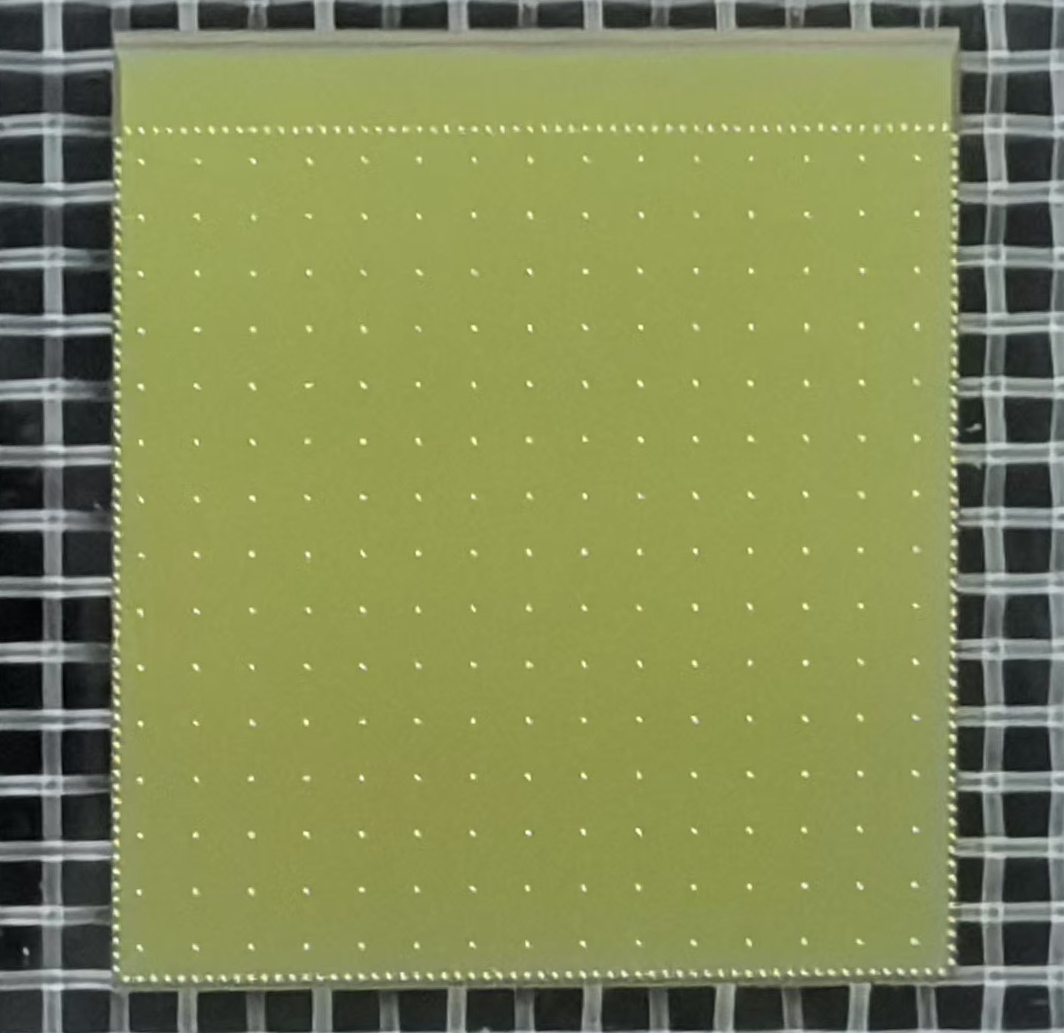}
        \caption{ASIC, with 240 bumps (white spots) on the guard ring, used in Type B240 hybrids.}
        \label{altiroca}
    \end{figure}

\begin{figure}[!h]
        \centering
        \includegraphics[width=0.5\linewidth]{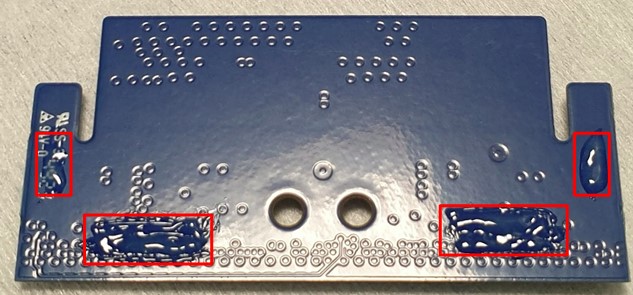}
        \caption{Glue dispensed on PCB for pattern 2. The measured total glue weight is 6.1~mg, similar to the estimated 6.6~mg in the 3-D model. }
        \label{jpattern}
    \end{figure}

Additional validation testing for the other proposed module designs is now in progress. The qualitiative agreement between simulation predictions and the experimental results from the initial test cases has validated the model's predictive power.  
Concurrently, the enhanced guard ring bump design has advanced to the fabrication phase (Figure~\ref{altiroca}), and the alternative adhesive Pattern 2 is undergoing experimental validation (Figure~\ref{jpattern}). Preliminary assessments of wire-bonded modules utilizing Pattern 2 indicate no observable issues; however, final production approval remains dependent on the results of upcoming thermal cycling tests.
    
\section{Conclusion}

This study evaluates the thermomechanical reliability of flip-chip bump bonds in HGTD modules under thermal cycling conditions, addressing challenges associated with bump disconnections. Using finite element simulations of modules subjected to thermal cycling, lifetime predictions were generated and several design improvements were proposed. The simulations indicate that implementing thicker sensors, increasing the peripheral bump density on the hybrids, and optimizing the adhesive application pattern can significantly extend module lifetime.

Initial thermal cycling tests of 775-\SI{}{\micro\meter} sensor modules (B30) show qualitiative agreement with simulated lifetime predictions and substantial improvement on prior 300-\SI{}{\micro\meter} sensor designs (A30), both with non-increased peripheral bumps (30 bumps per hybrid). Notably, the hybrid design incorporating an increased peripheral bump density (B240)--which is projected to substantially outperform the already-qualified B30 modules--was formally approved in the final design review for the HGTD modules and cleared for mass production. Although verification testing of additional design variants remains ongoing, these combined simulation and experimental results confirm that the optimized HGTD module designs are expected to meet the required robustness against temperature variations throughout the operational lifetime of the detector.

While the simulation results show qualitative agreement with experimental tests, further large-scale production testing is necessary to comprehensively validate the simulation methodology. These optimized module designs enable subsequent detector-level integration studies involving multiple assembled modules.

\acknowledgments

We gratefully acknowledge the financial support from the Slovenian Research and Innovation Agency (ARIS P1-0135, ARIS Z1-50011, ARIS J7-4419), Slovenia; the National Natural Science Foundation of China (No. 11961141014, No. 12188102) and the Ministry of Science and Technology of China (No. 2023YFA1605901), China; and FAPESP (2020/04867-2, 2022/14150-3, 2023/18486-9 and 2023/18484-6), MCTI/CNPq (INCT CERN Brasil 406672/2022-9) and CAPES - Code 001, Brazil. This work was partially funded by grant "MCIN Spain (PID2021-124660OB-C21) and by the Generalitat de Catalunya (AGAUR 2021-SGR-01506)". We acknowledge the support from BMFTR, Germany; NWO, Netherlands; and FCT, Portugal.

\bibliographystyle{JHEP}
\bibliography{paper}

\providecommand{\href}[2]{#2}\begingroup\raggedright\begin{thebibliography}{10}

\bibitem{particle_detector}
S.~Waid, J.~Maier, P.~Gaggl, A.~Gsponer, P.~Sieberer, M.~Babeluk et~al., \emph{Detector development for particle physics}, \href{https://doi.org/10.1007/s00502-023-01201-w}{\emph{e+i Elektrotechnik und Informationstechnik} {\bfseries 141} (2024) 20}.

\bibitem{CMS}
{\scshape CMS} collaboration, \emph{{The Phase-2 Upgrade of the CMS Tracker}},  Tech. Rep. CERN-LHCC-2017-009, CMS-TDR-014, CERN, Geneva (2017), \href{https://doi.org/10.17181/CERN.QZ28.FLHW}{DOI}.

\bibitem{ITk-TDR}
{\scshape ATLAS} collaboration, \emph{{Technical Design Report for the ATLAS Inner Tracker Pixel Detector}},  Tech. Rep. CERN-LHCC-2017-021, ATLAS-TDR-030, CERN, Geneva (2017), \href{https://doi.org/10.17181/CERN.FOZZ.ZP3Q}{DOI}.

\bibitem{strip}
{\scshape ATLAS} collaboration, \emph{{Technical Design Report for the ATLAS Inner Tracker Strip Detector}},  Tech. Rep. \href{https://cds.cern.ch/record/2257755}{CERN-LHCC-2017-005, ATLAS-TDR-025}, CERN, Geneva (2017).

\bibitem{belle}
T.~Abe, I.~Adachi, K.~Adamczyk, S.~Ahn, H.~Aihara, K.~Akai et~al., \emph{Belle ii technical design report},  2010.

\bibitem{TDR}
{\scshape ATLAS} collaboration, \emph{{Technical Design Report: A High-Granularity Timing Detector for the ATLAS Phase-II Upgrade}},  Tech. Rep. \href{https://cds.cern.ch/record/2719855}{CERN-LHCC-2020-007, ATLAS-TDR-031}, CERN, Geneva (2020).

\bibitem{MIP}
{\scshape CMS} collaboration, \emph{{A MIP Timing Detector for the CMS Phase-2 Upgrade}},  Tech. Rep. \href{https://cds.cern.ch/record/2667167}{CERN-LHCC-2019-003, CMS-TDR-020}, CERN, Geneva (2019).

\bibitem{flip-chip1}
A.~Yeo, C.~Lee and J.~Pang, \emph{Flip chip solder joint fatigue analysis using 2d and 3d fe models},  in \emph{5th International Conference on Thermal and Mechanical Simulation and Experiments in Microelectronics and Microsystems, 2004. EuroSimE 2004. Proceedings of the}, pp.~549--555, 2004, \href{https://doi.org/10.1109/ESIME.2004.1304090}{DOI}.

\bibitem{flip-chip2}
M.~Chen, Y.~Zhang, Y.~Guo, T.~Wu, L.~Dou, W.~Tian et~al., \emph{Research on flip-chip bonding process and thermal cycle reliability simulation of 3-d stacked structure}, \href{https://doi.org/10.1109/TCPMT.2021.3127157}{\emph{IEEE Transactions on Components, Packaging and Manufacturing Technology} {\bfseries 12} (2022) 51}.

\bibitem{viscomodel}
P.~Perzyna, \emph{Thermodynamic theory of viscoplasticity},  vol.~11 of \emph{Advances in Applied Mechanics}, pp.~313--354, Elsevier (1971), \href{https://doi.org/https://doi.org/10.1016/S0065-2156(08)70345-4}{DOI}.

\bibitem{visco}
G.~Xia, F.~Qin, W.~Zhu, C.~Gao and X.~Ma, \emph{Effects of solder constitutive models and fe models on fatigue life of dual-row qfn package},  in \emph{2012 13th International Conference on Electronic Packaging Technology \& High Density Packaging}, pp.~595--599, 2012, \href{https://doi.org/10.1109/ICEPT-HDP.2012.6474689}{DOI}.

\bibitem{anandmodel}
X.~Chen, G.~Chen and M.~Sakane, \emph{Prediction of stress-strain relationship with an improved anand constitutive model for lead-free solder sn-3.5ag}, \href{https://doi.org/10.1109/TCAPT.2004.843157}{\emph{IEEE Transactions on Components and Packaging Technologies} {\bfseries 28} (2005) 111}.

\bibitem{ansys}
{ANSYS, Inc.}, \emph{Ansys {Mechanical}},  2022.

\bibitem{anand}
L.~Jiang, W.~Zhu and H.~He, \emph{Comparison of darveaux model and coffin-manson model for fatigue life prediction of bga solder joints},  in \emph{2017 18th International Conference on Electronic Packaging Technology (ICEPT)}, pp.~1474--1477, 2017, \href{https://doi.org/10.1109/ICEPT.2017.8046714}{DOI}.

\bibitem{coffin-manson_use1}
P.~Sun, C.~Andersson, Z.~Cheng, Z.~Lai, D.~Shangguan and J.~Liu, \emph{Coffin-manson equation determination for sn-zn based lead-free solder joints},  in \emph{2005 Conference on High Density Microsystem Design and Packaging and Component Failure Analysis}, pp.~1--6, 2005, \href{https://doi.org/10.1109/HDP.2005.251421}{DOI}.

\bibitem{coffin-manson_use2}
S.~Jacques, A.~Caldeira, N.~Batut, A.~Schellmanns, R.~Leroy and L.~Gonthier, \emph{A coffin-manson model to predict the triac solder joints fatigue during power cycling},  in \emph{Proceedings of the 2011 14th European Conference on Power Electronics and Applications}, pp.~1--8, 2011.

\bibitem{coffin-manson1}
W.~Engelmaier, \emph{Fatigue life of leadless chip carrier solder joints during power cycling}, \href{https://doi.org/10.1109/TCHMT.1983.1136183}{\emph{IEEE Transactions on Components, Hybrids, and Manufacturing Technology} {\bfseries 6} (1983) 232}.

\bibitem{coffin-manson2}
K.~Wang, P.~Songling, Z.~Wangcheng, S.~Lin and D.~Chen, \emph{Thermal fatigue reliability analysis of charging module based on coffin-manson model and monte carlo},  in \emph{2023 Prognostics and Health Management Conference (PHM)}, pp.~55--58, 2023, \href{https://doi.org/10.1109/PHM58589.2023.00018}{DOI}.

\end{thebibliography}\endgroup
\end{document}